\documentclass[aps,prb,twocolumn,showpacs,superscriptaddress,floatfix]{revtex4-1}
\usepackage{bm}
\usepackage{amssymb}
\usepackage{graphicx}
\usepackage{amsmath}
\usepackage{dcolumn}
\usepackage[pdftex]{epsfig}
\usepackage{color}
\usepackage[german,english]{babel}
\usepackage{psfrag}
\usepackage{hyperref}
\usepackage{color}

\newcommand{\ket}[1]{\ensuremath{|#1\rangle }}
\newcommand{\hpr}[2]{\ensuremath{\langle #1|#2\rangle}}

\newcommand{\beq}{\begin{eqnarray}}
\newcommand{\eeq}{\end{eqnarray}}
\newcommand{\beqa}{\begin{equation}}
\newcommand{\eeqa}{\end{equation}}
\DeclareMathOperator{\Tr}{Tr}
\newcommand{\expec}[1]{\langle#1\rangle}

\newcommand{\eV}{\ \mathrm{eV}}
\newcommand{\meV}{\ \mathrm{meV}}

\begin{document}

\title{Variational exact diagonalization method for Anderson impurity models}
\author{M. Sch\"uler}
\email{mschueler@itp.uni-bremen.de}
\author{C. Renk}
\author{T. O. Wehling}
\affiliation{Institut f{\"u}r Theoretische Physik, Universit{\"a}t Bremen, Otto-Hahn-Allee 1, 28359 Bremen, Germany}
\affiliation{Bremen Center for Computational Materials Science, Universit{\"a}t Bremen, Am Fallturm 1a, 28359 Bremen, Germany}

\date{\today}

\begin{abstract}
We describe a variational approach to solving Anderson impurity models by means of exact diagonalization. Optimized parameters of a discretized auxiliary model are obtained on the basis of the Peierls-Feynman-Bogoliubov principle. Thereby, the variational approach resolves ambiguities related with the bath discretization, which is generally necessary to make Anderson impurity models tractable by exact diagonalization. The choice of variational degrees of freedom made here allows systematic improvements of total energies over mean field decouplings like Hartree-Fock. Furthermore, our approach allows us to embed arbitrary bath discretization schemes in total energy calculations and to systematically optimize and improve on traditional routes to the discretization problem such as fitting of hybridization functions on Matsubara frequencies. Benchmarks in terms of a single orbital Anderson model demonstrate that the variational exact diagonalization method accurately reproduces free energies as well as several single- and two-particle observables obtained from an exact solution. Finally, we demonstrate the applicability of the variational exact diagonalization approach to realistic five orbital problems with the example system of Co impurities in bulk Cu and compare to continuous-time Monte Carlo calculations. The accuracy of established bath discretization schemes is assessed in the framework of the variational approach introduced here.
\end{abstract}

\maketitle
\section{Introduction}
The Anderson impurity model\cite{anderson_localized_1961} (AIM) is a general model for the description of interacting impurities in metallic host systems. Originally, it was developed to describe single atoms with open d- or f-shells embedded in bulk materials and to understand the formation of their magnetic moments\cite{anderson_localized_1961}. Furthermore the model includes the widely discussed Kondo physics \cite{PhysRevB.21.1003}. Multi orbital variants of the AIM gained considerable attention in the context of rare-earth impurity systems\cite{PhysRevB.13.2950,PhysRevB.28.4315} as well as more recently magnetic adatoms or molecules on surfaces\cite{PhysRevLett.103.016803,PhysRevB.81.115427,surer_multiorbital_2012,Kue2014}. Finally, dynamical mean field theory\cite{georges_dynamical_1996} (DMFT) links correlated bulk systems as well as nanostructures to Anderson impurity models. 

To address the electronic structure of realistic correlated electron materials one often resorts to LDA++ approaches\cite{PhysRevB.57.6884}, where quantum lattice or impurity models are derived from first principles calculations. The resulting models are typically multi-orbital models including complex hybridization between the impurity and a continuous bath of states from the surrounding material, which brings along two challenges: First, the numerical solution of the impurity models and second the interpretation of the physics contained in these generally complex models in more simple terms. Experiments are for instance often interpreted in terms of atomic spins, crystal field, ligand field\cite{ballhausen1962ligand} or cluster approaches\cite{groot1994}, which typically involve a small discrete set of bath states or no bath states at all. The link of the complex, ab initio derived models and simpler phenomenological models is a priori unclear and relates to the so-called bath discretization problem of exact diagonalization solvers of the AIM.

The solution of the Anderson impurity model for general parameters has to be done numerically by means of e.g. quantum Monte Carlo\cite{CTQMC_RMP} (QMC), numerical renormalization group\cite{bulla_numerical_2007} (NRG), or exact diagonalization (ED) methods.\cite{georges_dynamical_1996} While NRG and QMC are in principle numerically exact methods, they become computationally very demanding, when dealing with many orbitals, hybridization functions with low symmetry, spin orbit coupling and general fermionic four operator Coulomb vertices.  ED methods deal with low symmetries and general Coulomb vertices at no additional computational cost but suffer from the so-called bath discretization problem: Due to the exponential growth of the many particle Fock space with the system size, it can handle only a few bath levels per orbital. A mapping of the continuous bath to a discrete version has to be found. Several approaches to this task have been introduced. One is to fit the hybridization function of the continuous bath on Matsubara frequencies \cite{caffarel_exact_1994}, another is to represent the hybridization function by a continued fraction and to link its coefficients to the parameters of the bath \cite{si_correlation_1994}.
These schemes are systematic in the sense that they converge to the full model when including more and more bath sites. However, in the multi-orbital case, the number of bath sites is limited (typically on the order of three or less for a five orbital impurity problem), so the quality of the mapping can hardly be checked by an analysis of the convergence.

Basically two different strategies have been laid out to circumvent this problem. First, the many body Hilbert space can be truncated in the sense of configuration interaction (CI) expansions, which have a long tradition in the context of quantum impurity problems\cite{PhysRevB.13.2950,PhysRevB.28.4315} and are subject of recent developments.\cite{zgid_truncated_2012,PhysRevB.90.085102,PhysRevB.90.235122} CI expansions are variational, i.e. they deliver upper bounds for total energies, but they do not provide simplified auxiliary Hamiltonians. On the other hand, there are several approaches towards optimized cluster approximations to Anderson impurity problems. In this context, self-energy functional theory\cite{PhysRevLett.91.206402} is based on an extremal principle but it is not variational regarding total energies and does not allow for variations of interaction parameters or the interacting orbitals. More general optimizations are possible in the framework of the so-called self-energy embedding theory (SEET)\cite{kananenka_systematically_2014}, which is however not variational.

In this paper, we combine ideas of variational approaches and optimized cluster approximations to the AIM. We introduce a strictly variational method of approximating an AIM with continuous bath by an AIM with finite strongly reduced number of bath sites, which we call variational ED method. It guaranties an optimal approximation to the AIM for a given number of bath sites in the sense of thermodynamic ground state properties. The method is based on the well-known Peierls-Feynman-Bogoliubov variational principle \cite{Peierls_1938, Bogoliubov_1958, Feynman_1972}, which finds optimal effective models on the basis of an optimal density matrix by minimizing a free energy functional. 

We will introduce the AIM and the variational principle in Sec. \ref{sec:meth}, where we also explain the details of calculating the Peierls-Feynmann-Bogoluibov
free energy functional and how to minimize it efficiently. By treating a single orbital model with the variational ED method in Sec. \ref{sec:bench} we analyze its performance in comparison to an exact treatment, established bath discretization methods\cite{caffarel_exact_1994} as well as Hartree-Fock theory. In Sec. \ref{sec:appl}, we demonstrate the applicability of the method to realistic five orbital system with the example of Co impurities in bulk Cu and compare to QMC simulations. We show that the variational ED method leads to systematically lower, i.e. more accurate, free energy estimates than unrestricted Hartree-Fock and traditional bath discretization schemes also in the multiorbtial case. Conclusions and outlook are given in Sec. \ref{sec:concl}.

\section{Model and method}
\label{sec:meth}
After introducing the Anderson impurity model we will recapitulate the Peierls-Feynman-Bogoliubov variational principle and show how to apply it to discretize Anderson impurity models in an optimal manner.

\subsection{The Anderson impurity model} The Hamiltonian of the initial AIM (termed ``original model'' hereafter) reads
\begin{align}
H = H_\text{bath} + H_\text{hyb} + H_\text{imp}. \label{eq:HSIAM}
\end{align}
The bath is described by
\begin{align}
H_\text{bath} = \sum_{\alpha k,\sigma} \varepsilon_{\alpha k} n_{\alpha k\sigma}^c\label{eq:HSIAM1},
\end{align} 
where $\varepsilon_{\alpha k}$ is the energy of the bath state with band/orbital index $\alpha$ and some additional quantum number $k$. $n_{\alpha k\sigma}^c=c^\dagger_{\alpha k\sigma}c_{\alpha k\sigma}$ is the corresponding particle number operator. The hybridization part
\begin{align}
H_\text{hyb} = \sum_{\alpha k,\sigma} V_{\alpha k} \left( c^\dagger_{\alpha k\sigma} d_{\alpha\sigma}
+ d^\dagger_{\alpha\sigma} c_{\alpha k\sigma}\right)\label{eq:HSIAM2}
\end{align}
couples the bath sites of one band to an orbital of the impurity with a coupling strength $V_{\alpha k}$. The bath electrons with spin $\sigma$ are created and annihilated by $c^\dagger_{\alpha k\sigma}$ and $c_{\alpha k\sigma}$, respectively, while $d^\dagger_{\alpha\sigma} $ ($d_{\alpha\sigma}$) denote the creation (annihilation) operators of the impurity electrons. The impurity site is described by
\begin{align}
H_\text{imp}=\sum_{\alpha,\sigma} \varepsilon_{\alpha}^d n_{\alpha\sigma}^d + \sum_{\alpha,\beta,\gamma,\delta,\sigma\sigma'} U_{\alpha \beta \gamma \delta} d^\dagger_{\alpha\sigma} d^\dagger_{\beta\sigma'} d_{\gamma\sigma'} d_{\delta\sigma}, \label{eq:origAIMLocal}
\end{align}
which contains the on-site Coulomb interaction $U_{\alpha \beta \gamma \delta}$ and the on-site energies $\varepsilon^d_{\alpha}$. By integrating out all bath degrees of freedom we arrive at the
hybridization function
\begin{align}
\Delta_\alpha(\omega) = \sum_k \frac{V_{\alpha k}^*V_{\alpha
k}}{\omega+i0^+-\varepsilon_{\alpha k}},
\end{align}
which describes the energy dependent coupling of the impurity to the bath.

\subsection{Peierls-Feynman-Bogoliubov variational principle}
Given a Hamiltonian $H$, which is ``difficult'' to solve, we search for an optimal approximation to $H$ within a set of simpler effective Hamiltonians $\tilde H$. The Peierls-Feynman-Bogoliubov variational principle\cite{Peierls_1938, Bogoliubov_1958, Feynman_1972} provides us with a prescription on how to fix the parameters of $\tilde H$ in a thermodynamically optimal way, i.e. such that the canonical density matrix resulting form $\tilde H$ approximates the density matrix corresponding to $H$ as close as possible. More strictly speaking: the canonical density operator $\rho_{\tilde H}=1/Z_{\tilde H} \exp(-\beta \tilde H)$ of the auxiliary system, where $Z_{\tilde H}=\Tr  \exp(-\beta \tilde H) $ is the partition function, approximates the exact density operator $\rho$ derived from $H$ as close as possible, when the Peierls-Bogoliubov-Feynman functional 
\begin{align}
\tilde{\Phi}[\rho_{\tilde H}] = \Phi_{\tilde H} + \expec{H-\tilde H}_{\tilde H},\label{eq:EDstar_functional}
\end{align}
becomes minimal. Here $\Phi_{\tilde H} = -\frac{1}{\beta} \ln Z_{\tilde H}$ is the free energy of the effective system. $\expec{H-\tilde H}_{\tilde H} =\Tr \rho_{\tilde H} (H-\tilde H)$ denotes a thermodynamic expectation value with respect to the effective system. In the case of $\rho_{\tilde H}=\rho$ the functional $\tilde{\Phi}[\rho_{\tilde H}]$ becomes minimal and coincides with the free energy $\Phi_H$ of the original system. In our case $H$ represents the full AIM, Eqs. (\ref{eq:HSIAM})-(\ref{eq:origAIMLocal}), and $\tilde H$ is the model with discretized bath, which is now introduced.

\subsection{Effective Hamiltonian}
\begin{figure}
\begin{center}
\mbox{
\includegraphics[width=1\columnwidth]{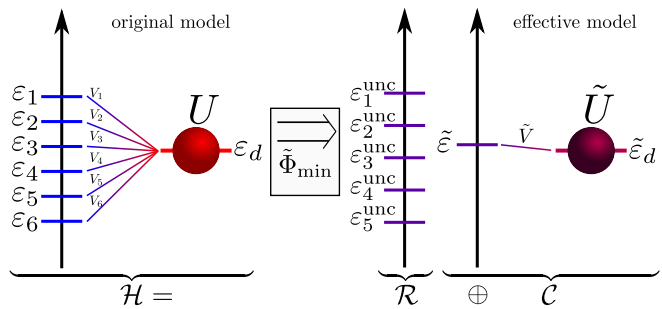}
}
\end{center}
\caption{(Color online) Illustration of the original and effective model for the case of one orbital and six bath sites. Blue represents bath character and red impurity character: In the effective model bath and impurity states can be mixed. $\varepsilon_n^\text{unc}$ are eigenvalues of $h^{\mathcal R}_{kk'}$.}
\label{fig:benchBath}
\end{figure}

The structure of the effective Hamiltonian for the case of a single impurity orbital is depicted in the right panel of Fig. \ref{fig:benchBath}. In contrast to the original model (left panel of Fig. \ref{fig:benchBath}), the effective model consists of two decoupled parts: First, the effective impurity coupled to one bath site only and second the remaining bath sites. I.e. we partition the full Hilbert space $\mathcal H$ into a correlated subspace $\mathcal C$ (first part) and an uncorrelated rest $\mathcal R$ (second part). In this work, we consider for concreteness a cluster consisting of a multi-orbital impurity and one bath site per impurity orbital for the correlated space but other choices are similarly possible. The single particle states of the effective model are related to those of the original model by a unitary transformation, which allows for mixing of original ``bath'' and ``impurity'' character in the effective model.

The optimal matrix elements of the effective model, as well as the optimal unitary transformation are found by minimizing the functional (\ref{eq:EDstar_functional}).
The states spanning $\mathcal C$ are defined by
\begin{align}
\ket{\tilde d_\alpha} &= u^{d_\alpha}_{d_\alpha} \ket{d_\alpha} + \sum_k u^{d_\alpha}_{c_{\alpha k}} \ket{c_{\alpha k}}, \label{eq:states_corrd} \\
\ket{\tilde c_{\alpha 1}} &= u^{c_{\alpha 1}}_{d_\alpha} \ket{d_\alpha} + \sum_k u^{c_{\alpha 1}}_{c_{\alpha k}} \ket{c_{\alpha k}},
\label{eq:states_corrc}
\end{align}
where the coefficients $u$ are chosen such that $\ket{\tilde d_\alpha}$ and $\ket{\tilde c_{\alpha 1}}$ form an orthonormal basis of $\mathcal{C}$. An orthonormal basis spanning $\mathcal R$ is defined by
\begin{align}
\ket{\tilde c_{\alpha k}} =u^{c_{\alpha k}}_{d_\alpha} \ket{d_\alpha} + \sum_{k'} u^{c_{\alpha k}}_{c_{\alpha k'}} \ket{c_{\alpha k'}}, k>1.
\label{eq:states_uncorr}
\end{align}
As a whole, the coefficients $u$ form a unitary matrix. In practice, we obtain the elements of this matrix from the QR decomposition of a matrix, in which the first two rows are defined by the coefficients of $\ket{\tilde d_\alpha}$ and  $\ket{\tilde c_{\alpha 1}}$ and all other elements are zero. This leads to a new orthonormal basis for the full space $\mathcal H$, which provides the partitioning according to $\mathcal H=\mathcal{C}\oplus\mathcal{R}$. The ansatz for the effective Hamiltonian in this new basis explicitly reads
\begin{align}
\tilde H = \tilde H ^\mathcal{C} + \tilde H ^\mathcal{R},
\end{align}
with
\begin{align}
\begin{split}
\tilde H^{\mathcal C} = \sum_\alpha \tilde V_\alpha \left( \tilde c^\dagger_{\alpha 1} \tilde d_\alpha + \tilde d^\dagger_\alpha \tilde c_{\alpha 1} \right) +  \sum_\alpha  \tilde \varepsilon_{\alpha 1} \tilde c^\dagger_{\alpha 1} \tilde  c_{\alpha 1} \\
+ \sum_\alpha \tilde \varepsilon_\alpha^d \tilde d^\dagger_\alpha \tilde d_\alpha+ \sum_{\alpha\beta\gamma\delta,\sigma\sigma'} \tilde  U_{\alpha\beta\gamma\delta}  \tilde d^\dagger_{\alpha\sigma}\tilde  d^\dagger_{\beta\sigma'}\tilde d_{\gamma\sigma'} \tilde d_{\delta\sigma} \label{eq:effHamilCorr}
\end{split}
\end{align}
and
\begin{align}
\tilde H^{\mathcal R} = \sum_{\alpha,(k,k')>1} h^{\mathcal R}_{\alpha kk'} \tilde c^\dagger_{\alpha k} \tilde  c_{\alpha k'}. \label{eq:effHamilUnCorr}
\end{align}

It is stressed, that the new states are linear combinations of the original impurity \textit{and} bath states, leading to mixed basis states. The new impurity states can have some amount of bath character and vice versa. The Hamiltonian in Eq. (\ref{eq:effHamilCorr}) states a many-body problem which can be solved by exact diagonalization, as long as its Hilbert space is sufficiently small. In contrast, the Hamiltonian
(\ref{eq:effHamilUnCorr}) states a one-particle problem and can be solved by diagonalizing the matrix $h^{\mathcal R}_{\alpha kk'}$. In summary, the Hamiltonian $\tilde H = \tilde H^\mathcal{C} + \tilde H^\mathcal{R}$ defines an effective Hamiltonian, which can be solved exactly and thus the functional (\ref{eq:EDstar_functional}) can be calculated. 

This ansatz implies several approximations. First, all couplings between $\mathcal{C}$ and $\mathcal{R}$ are neglected. Second interaction terms are restricted to new effective impurity orbitals $\tilde d_\alpha$ within $\mathcal{C}$, which is motivated by the fact that the original model includes only on-site interactions too. The latter approximation can be relaxed to include arbitrary interactions within $\mathcal{C}$, but we keep it here for simplicity.

Finally, we note that the amount of variational degrees of freedom in the variational ED approach is such that it includes Hartree-Fock as the limiting case $\tilde  U_{\alpha\beta\gamma\delta} \to 0$. Thus, we expect that variational ED will generally give more accurate energy estimates than Hartree-Fock.

\subsection{Implementation}

In order to perform the minimization in practice, the number of free parameters has to be kept sufficiently low. First, for the rest of this work it is assumed that the Coulomb tensor $\tilde U_{\alpha \beta\gamma\delta}$ is not varied. We choose it to be the same as in the original model. Test calculations have shown, that the variation of the Coulomb tensor is not crucial, as this can mostly be absorbed into the variation of the impurity level. The single particle matrix elements of $\tilde H^{\mathcal C}$ are assumed to be free parameters. In principle, the parameters of the uncorrelated Hamiltonian are free parameters, too. 
However, to further reduce the number of free parameters, we define $h^{\mathcal R}_{\alpha kk'}$ by a projection of a Hartree-Fock solution of the original Hamiltonian onto the states $\ket{\tilde c_{\alpha k}}$. The Hartree-Fock solution of the original Hamiltonian (\ref{eq:HSIAM}) can be written as 
\begin{align}
    H_\text{HF} = \sum_n \varepsilon^\text{HF}_n c^\dagger_n c_n,
\end{align}
where the eigenstates $\ket{n}$ and energies $\varepsilon^\text{HF}_n$ are found by applying the Hartree-Fock decoupling
\begin{align}
\begin{split}
d^\dagger_{\alpha\sigma} d^\dagger_{\beta\sigma'} d_{\gamma\sigma'} d_{\delta\sigma} \rightarrow & \langle d^\dagger_{\alpha\sigma} d_{\delta\sigma} \rangle  d^\dagger_{\beta\sigma'} d_{\gamma\sigma'}  + \langle d^\dagger_{\beta\sigma'} d_{\gamma\sigma'} \rangle d^\dagger_{\alpha\sigma} d_{\delta\sigma} \\
 - &\langle d^\dagger_{\alpha\sigma}  d_{\gamma\sigma'} \rangle d^\dagger_{\beta\sigma'} d_{\delta\sigma}- \langle d^\dagger_{\beta\sigma'} d_{\delta\sigma} \rangle d^\dagger_{\alpha\sigma}  d_{\gamma\sigma'}
\end{split}
\end{align}
to (\ref{eq:origAIMLocal}) and solving the resulting non-interacting problem self-consistently. The single particle matrix elements within the uncorrelated space $\mathcal{R}$ explicitly read
\begin{align}
h^{\mathcal R}_{\alpha kk'} = \sum_n \varepsilon_n^\text{HF} \hpr{\tilde c_{\alpha k}}{n} \hpr{n}{\tilde c_{\alpha k'}}.
\end{align}
In order to not break any spin rotation symmetries, restricted Hartree-Fock is used.

The functional $\tilde \Phi[\rho_{\tilde H}]$ now depends on the unitary transformation and on the matrix elements of $\tilde H^\mathcal{C}$. The minimum of the functional is searched by iterative methods. Thus, the functional $\tilde \Phi[\rho_{\tilde H}]$ has to be calculated for various points of the variational space with the computationally most expensive part being here the diagonalizations of $\tilde H^\mathcal{C}$. Therefore, we first search for fixed parameters in $\tilde H^\mathcal{C}$ a corresponding optimal unitary transformation matrix defining the optimal partitioning $\mathcal H=\mathcal C \oplus \mathcal R$ using an SLSQP algorithm \footnote{A sequential least squares programming algorithm as implemented in the package scipy.optimize.minimize \cite{kraft1988software}}. The search of the minimum w.r.t. the parameters of $\tilde H^\mathcal{C}$ is then done by the Nelder-Mead algorithm \footnote{A simplex algorithm as implemented in the python module scipy.optimize.minimize\cite{nelder_simplex_1965}}. The number of independent parameters can be further reduced when the original system shows symmetries like orbital degeneracies which are assumed not to be broken in the effective model.

\section{Benchmark for a single orbital AIM}
\label{sec:bench}

In this section the variational ED method is tested for its performance in reproducing the density operator as well as observables such as the occupation number, double occupancy and crystal orbital overlap populations of a simple original model. The original model we consider here is a single orbital model with only 6 bath sites, which itself can be solved by exact diagonalization. The detailed setup of the model is as follows: The impurity level is $\varepsilon_d=-2.0\eV$, the interaction strength is $U=4.0\eV$. The 6 bath levels are equally aligned around a mean bath energy $\varepsilon_b$ in an interval of $2\eV$ (i.e. the bandwidth of the bath). The coupling is $V_k=0.9\eV$. The mean bath energy $\varepsilon_b$ is swept from $-6.0\eV$ to $6.0\eV$.  All energies are measured w.r.t. to the Fermi energy $\varepsilon_F=0$. The system is solved for $T=0$. The model is first solved exactly, second by the variational ED method, third by unrestricted Hartree-Fock and finally by ED using reduced bath sites obtained by fitting of the hybridization functions on the imaginary Matsubara frequency axis\cite{caffarel_exact_1994}. The latter type of approaches require generally the introduction of a so-called weight function $W_n$ for the fitting procedure, as explained in the appendix (\ref{app:imag}).

\begin{figure}
\mbox{
\includegraphics[width=1\columnwidth]{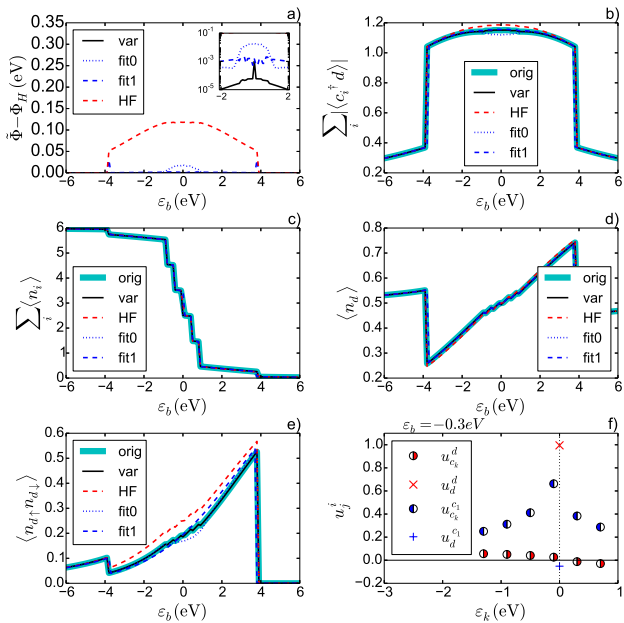}
}
\caption{(Color online) Benchmark of different ED approaches and spin-polarized Hartree-Fock theory against an exact solution for single orbital Anderson impurity models with a mean bath energy $\varepsilon_b$. (a) Difference between the free energy functional obtained by different approximate methods according to Eq. (\ref{eq:EDstar_functional}) and the free energy of the original model. The inset shows a close up view for the vicinity of the Fermi energy on a logarithmic scale. (b)-(e): Comparison of local and non-local observables obtained from an exact solution (``orig'', bold cyan) and calculated by the four different approximate methods, i.e. the variational ED method (``var'', solid black), Hartree-Fock (``HF'', dashed red) and fits of hybridization functions on the imaginary axis with different weight functions ($W_n=1$: ``fit0'', dotted blue; $W_n=1/\omega_n$: ``fit1'', dashed blue.) Panel b) shows the chemical bond strength, c) shows the total bath occupation, d) the impurity occupation and e) the double occupation. Panel e) shows the coefficients of the unitary transformation linking the original model to the optimized effective model with one bath-site per spin-orbital in $\mathcal{C}$ for the example of $\varepsilon_b = -0.3\eV$.}
\label{fig:benchRes}
\end{figure}

The central object for the assessment of the quality of the methods is the difference between $\tilde \Phi [\rho_{\tilde H}]$ and the exact free energy $ \Phi_H$, as shown in Fig. \ref{fig:benchRes}a). For bath sites energetically far away from the Fermi level and from single particle excitation energies of the impurity ($|\varepsilon_b|>4\eV$), all methods lead essentially to the correct free energy. Deviations occur, however, for bath levels closer to the Fermi energy. The Hartree-Fock free energy differs from the exact thermodynamical potential on the order of $100\meV$  basically in the whole range of $|\varepsilon_b|<4\eV$. The fitting of the hybridization function function on the imaginary axis leads to rather accurate free energies as long as all bath sites are above or below the Fermi energy ($|\varepsilon_b|>1\eV$), while for $|\varepsilon_b|<1\eV$ deviations from the exact thermodynamical potential on the order of $10$ to $20\meV$ occur. The choice of an optimal weight function (see appendix \ref{app:imag}) depends on details of the bath: For the case of bath sites on both sides of the Fermi level ($|\varepsilon_b|<1\eV$) $W_n=1/\omega_n$ leads to the lowest free energies. Otherwise, the constant weight function $W_n=1$ shows smallest deviations of the free energy functional from the exact solution. The variational ED method is generally very close to the exact solution. Only for the special case of a strictly symmetric distribution of the bath sites around the Fermi energy ($\varepsilon_b = 0\eV$) a deviation on the order of meV occurs.

Fig. \ref{fig:benchRes}b)-e) shows a comparison of several observables (chemical bond strength b), bath occupation c), impurity occupation d) and double occupation e)) calculated with the different methods. For the outer most regions ($|\varepsilon_b|>4\eV$) all methods describe the observables accurately. The fitting of hybridization functions on the Matsubara axis leads to deviations depending on the weight function, especially for the double occupation and chemical bond strength if the bath is centered around the Fermi energy ($\varepsilon_b\approx 0\eV$). Hartree-Fock systematically overestimates the double occupancy for $|\varepsilon_b|<4\eV$. The variational ED method shows nearly no deviations from the exact solution at all.

It is instructive to examine the unitary transformation linking the basis of the original and effective model. Fig. \ref{fig:benchRes}f) shows the coefficients of the linear combination of the states spanning the correlated space $\ket{\tilde d} $ and $\ket{\tilde c_1}$ (see Eqs. (\ref{eq:states_corrd}) and (\ref{eq:states_corrc})) for an original model with the bath centered around the energy $\varepsilon_b=-0.3\eV$. The effective impurity has mainly $\ket{d}$ character with small bath admixture and can approximately be interpreted as the old impurity state. The coupled effective bath state is nearly a pure linear combination of old bath states, where states closer to the Fermi energy contribute stronger than those further away. This behavior is very reminiscent of effective bath wave functions obtained in variational approaches like the Varma-Yafet\cite{PhysRevB.13.2950} or the Gunnarsson-Sch\"onhammer expansion.\cite{PhysRevB.28.4315}

For the treatment of original models with far more bath sites, it is important to note, that the coefficients defining the unitary transformation from the original bath states to the effective impurity and bath orbitals, i.e. $u^{c_1}_{c_k}$ and $u^{d}_{c_k}$, vary smoothly as function of the bath energies on either side of the Fermi energy.

\section{Co impurities in Cu: Application to a realistic five orbital system}
\label{sec:appl}
\subsection{The AIM derived from LDA}
In this section the variational ED method is applied to a realistic model of Co impurities in bulk Cu, which has been obtained from super-cell DFT calculations and has been analyzed using a QMC impurity solver in Ref. \onlinecite{surer_multiorbital_2012}. 

The cubic symmetry of the Cu crystal leads to a splitting of the Co $3d$-orbitals into blocks of t$_{2\text{g}}$ and e$_\text{g}$ symmetry. From the DFT hybridization function, which is a continuous function, we obtain our initial model assuming some large number of bath sites, here 100 per orbital. (This number does not present a limiting factor and could be chosen arbitrarily larger). The bath sites are assumed to be equidistantly distributed between $-10\eV$ and $10\eV$, and the hybridization terms $V_{\alpha k}$ are then found by fitting the imaginary part of a discretized hybridization function
\begin{align}
\Delta_\text{disc}(\omega) = \sum_k \frac{V_{\alpha k}^*V_{\alpha k}}{\omega - \varepsilon_k+i\delta},
\end{align}
with some broadening $\delta=0.1\eV$ to the ab initio hybridization function $\Delta(\omega)$ on the real axis. The $V_{ik}$ are plotted in Figure \ref{fig:CoCu}a). The crystal field obtained from the DFT calculation is $\varepsilon^{d}_{\text{e}_\text{g}}  - \varepsilon^{d}_{\text{t}_\text{2g}} = 0.136\eV$. As in Ref. \onlinecite{surer_multiorbital_2012}, we consider a rotationally invariant Coulomb interaction defined by
\begin{align}
U_{\alpha\beta\gamma\delta} = \sum_{k=0}^{2l} a_k(\alpha_m\beta_m,\gamma_m\delta_m)F^k,
\end{align}
where $a_k(\alpha_m\beta_m,\gamma_m\delta_m)$ are the Gaunt coefficients \cite{slater1960quantum,pavarini_correlated_2012:eder} and where $F^0 = U$, $F^2 = 14/(1+0.625)J$ and $F^4=0.625F^2$ are Slater parameters with the average Coulomb interaction $U=4.0\eV$ and Hunds exchange interaction $J=0.9\eV$. Due to the so-called double counting problem inherent to LDA++ approaches, the filling of the impurity d-levels is not exactly known. Here, we consider the double counting potential $\mu=27\eV$ as in Ref. \onlinecite{surer_multiorbital_2012}. All data is obtained at a inverse temperature of $\beta=40$, like in the case of the QMC simulations. Finally, we assume that the cubic symmetry of the system prevails, which means that only two independent sets of matrix elements (for the t$_\text{2g}$ and e$_\text{g}$ states) have to be varied during the minimization of $\tilde \Phi[\rho_{\tilde H}]$.

\subsection{Implementation of the variational ED method for the 5 orbital AIM}

We compare two different sets of variational degrees of freedom for the optimization of the one particle basis, which we refer to as ``bath'' and ``all''. In the  ``bath'' case, only bath sites are optimized, i.e. we fix the expansion coefficients $u_{d_\alpha}^{c_{\alpha 1}}=0$, $u_{c_{\alpha k}}^{d_{\alpha}}=0$ and $u_{d_\alpha}^{d_{\alpha}}=1$. This leads to considerably less variational parameters and a much smaller amount of expectation values to be calculated in each step of the iteration. In the second approach, ``all'', which is computationally more demanding because the full two-particle density matrix of the effective system has to be calculated, we optimize the full one particle basis of the bath and that of the impurity.

Because a full optimization of the parameters of the effective model is computationally challenging, it is crucial to start the optimization from a good initial guess. We obtain such initial guesses for the parameters of the bath by fitting of hybridization functions on Matsubara frequencies as introduced in the appendix \ref{app:imag}. We choose $\tilde\varepsilon^d_\alpha=  \varepsilon^d_\alpha$ as the initial guess for the parameters of the impurity. The resulting first guesses using different weight functions are summarized in the Table \ref{tab:CoCuCaf}. While all weight functions lead to setups with the t$_\text{2g}$ bath sites above the Fermi energy and the e$_\text{g}$ bath sites below, the details of their energetic positions and the hybridization strengths depend strongly on the form of $W_n$. Adding more weight on features on small Matsubara frequencies shifts the effective bath parameters to smaller values. The quality of these starting guesses in the context of the variational principle is discussed in the next section.

\begin{table}[htb]
\caption{Parameters of the effective model (see Eq. (\ref{eq:effHamilCorr})) obtained by the fit of hybridization functions on imaginary frequencies using different weight functions ($W_n=1,1/\omega_n,1/\omega_n^2$, see appendix \ref{app:imag}) and the iterative optimization (``var''). }
\begin{ruledtabular}
\begin{tabular}[b]{crrrr}
weight function &$1$& $1/\omega_n$&$1/\omega_n^2$&var \\ \colrule
$\tilde \varepsilon_{\text{t}_\text{2g}1}$(eV)& $ 3.203$ & $ 0.775$ & $ 0.068$ & $2.658$ \\
$\tilde V_{\text{t}_\text{2g}}$(eV)           & $ 1.563$ & $ 0.606$ & $ 0.223$ & $1.717$ \\
$\tilde \varepsilon_{\text{e}_\text{g}1}$(eV) & $-2.314$ & $-0.019$ & $-0.015$ & $-1.995$ \\
$\tilde V_{\text{e}_\text{g}}$(eV)            & $ 1.049$ & $ 0.170$ & $ 0.156$ & $1.418$ \\
$\tilde \varepsilon^d_{\text{t}_\text{2g}}$(eV) & $-27.30$ & $-27.30$ & $-27.30$ & $-26.57$\\
$\tilde \varepsilon^d_{\text{e}_\text{g}}$(eV) & $-27.44$ & $-27.44$ & $-27.44$ & $-27.87$

\end{tabular}
\label{tab:CoCuCaf}
\end{ruledtabular}
\end{table}

The large number of bath sites (100 per orbital) in the original model leads to 202 variational parameters defining the unitary transformation in the ``all'' case or 100 parameters in the ``bath'' case for each orbital. The observation, that $u^{d_\alpha}_{c_{\alpha k}}, u^{c_{\alpha 1}}_{c_{\alpha k}}$ are smooth functions of energy (c.f. Fig. \ref{fig:benchRes}e) below and above $E_F$, leads to the possibility of expanding them in a set of smooth functions and thereby reducing the number of variational parameters considerably. Here, we chose five Chebyshev polynomials $T_n(k)$ per orbital for bath sites above and five for those below the Fermi energy. Therefore only 22 (or 10 in the case of ``bath'') parameters per orbital have to be varied to find the optimal unitary transformation to embed the effective model into the full Hilbert space.

\subsection{Results}
We will first compare free energy estimates as well as different local observables obtained from variational ED treatments to unrestricted Hartree-Fock as well as QMC calculations. Afterwards, we investigate the nature of the optimized effective bath and impurity states as obtained from the variational ED treatment.

\subsubsection{Free energy functional and local observables}

\begin{table}[tb]
\caption{The free energy functional $\tilde \Phi$, the total impurity occupancy $n_d$ and the local spin $S$ as obtained from simulations of the AIM for Co impurities in Cu. The values of $\tilde \Phi$ are shown as differences to the results from unrestricted Hartree-Fock (UHF): $\Delta \tilde \Phi = \tilde \Phi - \tilde \Phi_\text{UHF}$. Total impurity occupation and spin calculated with the variational ED method are compared to QMC solutions of the AIM from Ref. \onlinecite{surer_multiorbital_2012}. Different flavors of the variational ED method are considered: first ``bath'' and second ``all'' with the model parameters obtained from the fits of the hybridization function on the imaginary frequencies using different weight functions ($W_n=1$ ``fit0'',$W_n=1/\omega_n$ ``fit1'' and $W_n=1/\omega_n^2$ ``fit2'') and finally full optimization of transformation and model parameters labeled ``var''. Restricted Hartree-Fock (RHF) and unrestricted HF (UHF) results are also shown.}
\begin{ruledtabular}
\begin{tabular}[b]{lccc}
 & $\Delta\tilde \Phi$(eV) & $\langle n_d\rangle$ & $S$ \\ \colrule
QMC       &	      & 7.78 $\pm$ 0.05&0.92 $\pm$ 0.02\\
UHF       & 0     & 7.78 & 1.78\\
RHF       & 0.52  & 8.20 & 1.06\\
fit0,bath & 0.05  & 7.75 & 1.06\\
fit1,bath & 1.15  & 7.84 & 1.04\\
fit2,bath & 2.77  & 7.92 & 1.03\\
fit0,all  & -0.22 & 7.71 & 1.03\\
fit1,all  & 0.32  & 7.65 & 1.05\\
fit2,all  & 1.04  & 7.60 & 1.00\\
var,bath  & 0.00  & 7.76 & 1.05\\
var,all   & -0.30 & 7.75 & 1.02\\

\end{tabular}
\label{tab:CoCu}
\end{ruledtabular}
\end{table}

Table \ref{tab:CoCu} shows the free energy functional $\tilde\Phi$ (relative to unrestricted Hartree-Fock (UHF)), as obtained with different starting points and different amounts of variational degrees of freedom in the variational ED approach. In the case of ``bath'', the constant weight function (``fit0'', $W_n=1$) leads to the lowest values of the $\tilde\Phi[\rho]$. The models derived using weight functions $W_n=1/\omega_n$ (``fit1'') and $W_n=1/\omega_n^2$ (``fit2'') lead to free energy estimates which are about $1$ to $3\eV$ higher in energy. The situation for the case of ``all'' is similar. On this basis, we have chosen the starting guess obtained with the constant weight function for the full optimization of the effective model parameters. The resulting parameters are shown in the last column of Tab. \ref{tab:CoCuCaf} and are close to the starting guess. The full optimization schemes (``var,bath'' and ``var,all'') find parameters which lower the functional $\tilde\Phi$ considerably for ``all'' and slightly for ``bath''.

Regarding the impurity occupation ($\langle n_d\rangle$, see Tab. \ref{tab:CoCu}), we see that the description by unrestricted Hartree-Fock is rather close to QMC, whereas restricted Hartree-Fock overestimates the occupation. All versions of exact diagonalization lead to occupations close to the QMC results, and many cases within the QMC error bars. The spin $S$ (defined as $\langle \hat{S}^2 \rangle=S(S+1)$), which is a two-particle observable, reveals the problems of the Hartree-Fock description. $S$ is vastly overestimated by unrestricted Hartree-Fock. The variational ED methods, especially the ``all'' case for the constant weight function (``fit0'') and the full optimized ED model, lead to results close to QMC.

To compare the results of the variational ED method with those ED methods based on fitting of the hybridization function on the imaginary axis, we should compare the ``fit0,bath'',``fit1,bath'', and ``fit2,bath'' cases to the corresponding ``all'' and ``var,all'' cases. We see that having more variational degrees of freedom leads to an improved description of the free energies, as it should be.

In general, we learn that only in the case of optimizing both effective bath and effective impurity states (termed ``all'') we reach lower values of the free energy functional $\tilde\Phi$ than with unrestricted Hartree-Fock: The freedom to form mixtures of bath and impurity states in the effective model is important to describe the free energy and local observables of the system adequately. As the variational ED method provides more accurate (free) energy estimates than unrestricted Hartree-Fock, the approach introduced here could be a way to improve LDA+U total energy schemes.

\subsubsection{The effective basis states}
\begin{figure}[h]
\mbox{
\includegraphics[width=0.99\columnwidth]{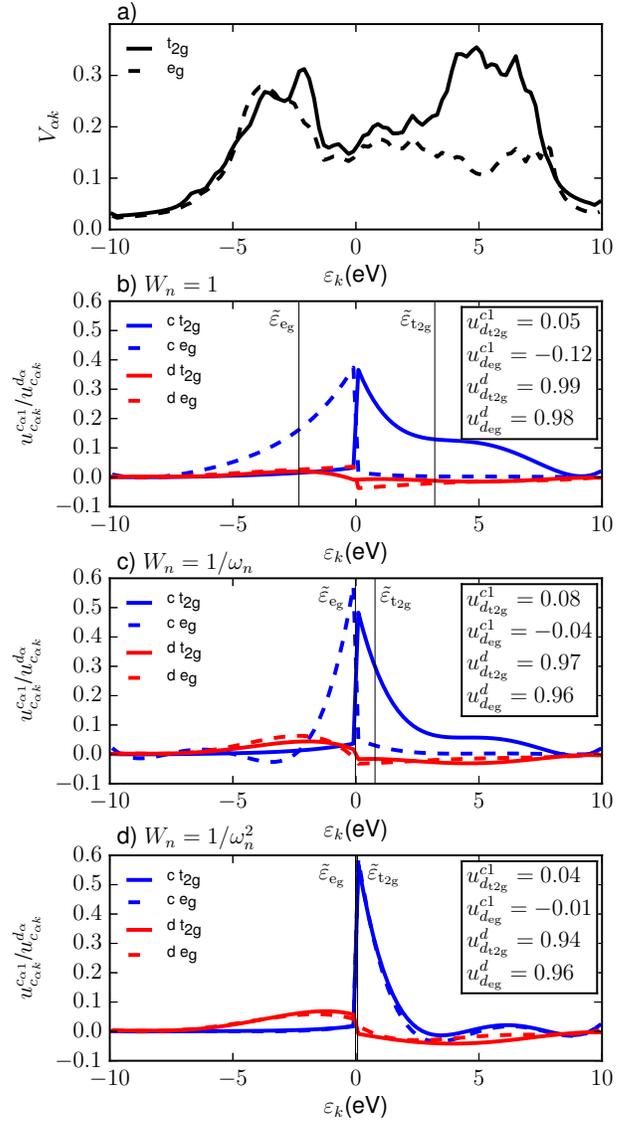}
}
\caption{(Color online) a) Hopping matrix elements between the impurity orbitals and the bath from the original AIM for Co impurities in Cu (solid t$_\text{2g}$, dashed e$_\text{g}$). b)-d) Coefficients defining the optimal transformation from original bath states to effective bath states (blue/dark gray) and effective impurity states (red/light gray), c.f. Eqs. (\ref{eq:states_corrd}) and (\ref{eq:states_corrc}). Optimized transformations for different effective models defined through fits of the hybridization with weight functions $W_n=1$ b), $W_n=1/\omega_n $ c), and $W_n=1/\omega_n^2$ d) are shown. The energies of the effective coupled bath sites $\tilde \varepsilon_{\alpha}$ are depicted as thin vertical lines. The numerical values of the transformation coefficients defining the admixture of original impurity states to the effective bath and impurity states are given as insets.}
\label{fig:CoCu}
\end{figure}

Now we analyze the unitary transformation relating the optimized basis states of the effective model and the original basis states. The transformation obtained for the models from the starting guesses with weight functions $W_n=1$, $1/\omega_n$, and $1/\omega_n^2$ is shown in Fig. \ref{fig:CoCu} b), c), and d), respectively. We observe a clear trend that the admixture of the original bath states into the effective bath state ($u_{c_{\alpha 1}}^{c_{\alpha k}}$, blue lines) is strongest in the vicinity of the effective bath site energy $\tilde\varepsilon_{\alpha1}$. For bath states close to the Fermi energy we get a sharp cut off for states on the opposite side of the Fermi energy. This is very similar to first order configuration interaction treatments of the AIM\cite{PhysRevB.28.4315,PhysRevB.13.2950}. The original impurity admixture in the effective impurity ($u^{d}_{d_\alpha}$) rises with the distance of the effective bath site from the Fermi energy. 

\section{Spectral functions}
The variational principle results in an effective model which represents thermodynamic ground state properties in an optimal manner. This is a necessary but not a sufficient condition to give a good approximation also for excitation spectra. In the following, we study the one particle spectral function for single orbital impurity benchmark systems from Sec. \ref{sec:bench} with the bath states centered around $\varepsilon_b=0.3\eV$ and two different hybridization strengths, $V_k=0.9 \eV$ and $V_k=0.3 \eV$, respectively. 
The impurity spectral function is obtained from the Lehmann representation of the impurity Green's function
\begin{align}
G_\alpha(\omega) =  \frac{1}{Z}  \sum_{\mu \nu}  \frac{\left| \langle \mu | d_\alpha^\dagger | \nu \rangle \right|^2}{\omega + E_\nu - E_\mu - i0^+} \left( e^{-\beta E_\nu} - e^{-\beta E_\mu}  \right), \label{eq:spec}
\end{align}
where in our calculations $0^+$ is replaced by a broadening of $\delta = 0.1\eV$ and the inverse temperature is $\beta=3200$, which is very close to the $T=0$ calculations of expectation values in Sec. \ref{sec:bench}. 

We assess the quality of the spectra obtained from the variational ED method, from ED with the hybridization function fitted on the Matsubara axis (with two different weight functions $W_n=1$ and $W_n = 1/ \omega_n$) and from an unrestricted Hartree-Fock treatment by comparing to the exact spectrum of the original model. For the case of $W_n = 1/ \omega_n$, we additionally compare the spectra for different amounts of variational freedom in choosing the basis states of the effective model, where we either optimized the bath states only (termed ``bath'', c.f. Sec. \ref{sec:appl}) or allowed impurity and bath states to mix (termed ``all''). 

\begin{figure}[h]
\mbox{
\includegraphics[width=0.99\columnwidth]{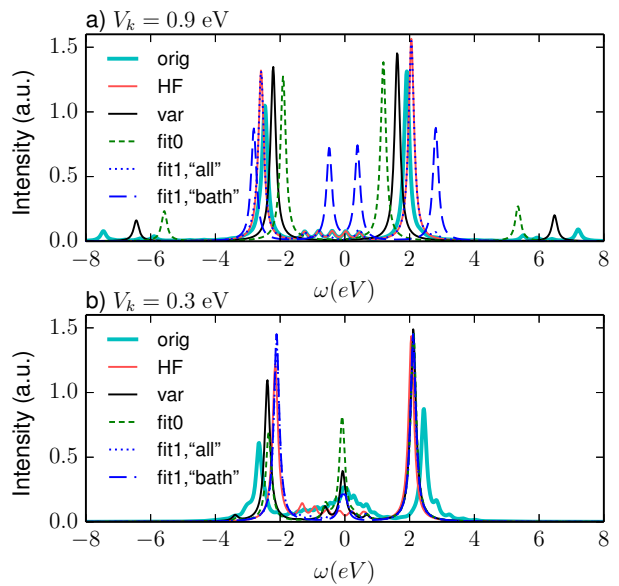}
}
\caption{(Color online) One particle impurity spectral functions for the single-orbital benchmark model introduced in Sec. \ref{sec:bench} with $\varepsilon_b=-0.3 \eV$ and hybridization strengths $V_k=0.9\eV$ (a) and $V_k=0.3\eV$ (b). The spectral function from the exact solution of the original model is shown in bold cyan, from the Hartree-Fock calculation in red and from the variational ED method in black. Spectra from ED with fitted hybridization functions on the Matsubara axis are depicted in dashed green (``fit0'', i.e. weight function $W_n=1$), dotted blue (``fit1'', $W_n=1/\omega_n$ optimizing all states) and dashed blue (``fit1'', $W_n=1/\omega_n$ optimizing only bath states).}
\label{fig:spectra}
\end{figure}

The dominant features of the original spectrum in the case of strong hybridization ($V_k=0.9\eV$, Fig. \ref{fig:spectra} a)) are two major peaks at about $-2.5\eV$ and $2\eV$ (stemming from bonding and anti-bonding combinations of impurity and bath orbitals), two satellite peaks far from the Fermi energy and additional smaller peaks around the Fermi energy. The spectral function from Hartree-Fock reproduces the bonding/anti bonding peaks and those close to the Fermi energy very well, while the satellites are missing. The variational ED method describes the positions of the main peaks well and also reproduces the satellite peaks, whereas the minor peaks around the Fermi energy are not present. The spectral function from a fit on the imaginary axis with a constant weight function ($W_n=1$) shows a similar picture but with major peaks and satellites shifted considerably towards the Fermi energy. The result for the weight function emphasizing small Matsubara frequencies ($W_n=1/\omega_n$) leads to a good representation of the peaks around $-2.5\eV$ and $2\eV$ and some minor peaks around the Fermi energy but the satellites are missing completely. The resulting spectrum obtained by not mixing bath and impurity states (``bath'') shows only little resemblance to the original spectrum. I.e. in this case the mixing of bath and impurity basis states can not only improve total energies / thermodynamic potentials but also spectra quite significantly. 

In the case of weaker hybridization ($V_k=0.3\eV$, Fig. \ref{fig:spectra} b)) the original impurity spectral function shows two Hubbard peaks at about $-2.6\eV$ and $2.4\eV$ and in comparison to the former case more spectral weight and additional features close to the Fermi energy. Here, the Hartree-Fock description results in a spin-polarized ground state and describes the positions of the Hubbard peaks in the spectrum correctly. The enhanced spectral weight at the Fermi energy is however not reproduced. Exact diagonalization of the effective models obtained from the variational method and from the fits of the hybridization functions on imaginary axis give similar results. In addition to the upper and lower Hubbard peaks the ED methods also reproduce enhanced spectral weight at the Fermi level.

While the performance of the fit methods in reproducing the spectra of the original model differs between the case with strong and weak hybridization  particularly for the weight function $W_n=1/\omega_n$, the variational method gives satisfactory results in both cases. The spectra from variational ED are in both cases at least as close to the spectra of the original model as the best spectrum obtained with any of the two bath fitting procedures $W_n=1$ or $1/\omega_n$) under investigation.

\section{Conclusion}
\label{sec:concl}
In conclusion, we present a variational exact diagonalization method which provides self-consistently optimized parameters of discretized Anderson impurity models. The method is based on an optimal partitioning of the system into a correlated part, where electronic interactions are explicitly taken into account, and an uncorrelated rest and on finding optimal effective Hamiltonians for both parts of the system. To this end, a variation of a free energy functional w.r.t. one-particle basis states spanning the correlated subspace and the matrix elements of the effective Hamiltonians is performed. 

A benchmark of the variational ED method against an exact solution of a one orbital Anderson model demonstrates its excellent performance in reproducing ground state observables of the impurity and bath and additionally a sound performance in reproducing the impurity spectral function. A comparison with Hartree-Fock and established bath discretization schemes for ED shows that the variational approach introduced here even works for difficult cases, i.e. when the bath is symmetric around the Fermi energy. Furthermore, applicability of the variational ED method to realistic multi-orbital cases is demonstrated with the example of Co impurities in bulk Cu. Also here, the variational method leads to an accurate description of local one and two particle observables like the impurity occupation and the spin. Energetically the method outperforms unrestricted Hartree-Fock, which suggests that the variational ED approach could be useful to improve total-energy approaches to correlated systems beyond LDA+U. Finally, the method introduced, here, can be used to embed established bath discretization schemes such as the fit of hybridization functions on Matsubara frequencies into a variational framework and to reach an unbiased decision of e.g. which weight function to choose in the fits of the hybridization functions. In the example studied, here, the constant weight function leads to best results in terms of the free energy, whereas $W_n=1/\omega_n^2$ leads to results qualitatively similar to a first order configuration interaction expansion\cite{PhysRevB.28.4315,PhysRevB.13.2950}. For the description of systems closer to the atomic limit, like Co on Cu or 4f systems we expect CI expansions to converge faster and weight functions like $W_n=1/\omega_n$ or $W_n=1/\omega_n^2$ might be a better choice. The presented variational approach will allow for an unbiased decision in any case.

Additionally, the variational method is universal and different choices of the effective correlated space are possible. To gain yet higher accuracy, more bath levels could be included. On the other hand, to make contact with ligand field theory or crystal field theory descriptions of magnetic impurity systems and nanostructures one could consider correlated subspaces with very few or without any effective bath orbitals at all.

\section{Acknowledgments}
The authors thank M. Katsnelson, A. Lichtenstein, M. Potthoff, P. Bl\"ochl, R. Schade ans S. Barthel for useful discussions as well as the Central Research Development Fund of the University of Bremen and the DFG via FOR 1346 for financial support.

\appendix
\section{Fit of hybridization functions on the Matsubara axis}
\label{app:imag}
The method of fitting hybridization functions is shortly introduced for the sake of completeness. The basic idea is to minimize a cost function for the inverse impurity Green's function (or equivalently the hybridization function) of the discretized model and that of the original model, both defined on the imaginary frequency axis \cite{caffarel_exact_1994}. In the case of one effective bath site, the discrete impurity Green's function is defined as
\begin{align}
g_0(i\omega_n) = \left( i\omega_n - \epsilon_d - \mu - \frac{\tilde V^2}{i\omega_n - \tilde \varepsilon_1}  \right)^{-1}
\end{align}
and the Green's function of the original model as
\begin{align}
G_0(i\omega_n) = \left( i\omega_n - \epsilon_d - \mu - \sum_{k} \frac{V_k^2}{i\omega_n - \varepsilon_k}  \right)^{-1}.
\end{align}
The cost function then reads
\begin{align}
\chi^2 = \frac{1}{n_\text{max} + 1} \sum_{n=0}^{n_\text{max}} W_n \left| G_0^{-1}(i\omega_n) - g_0^{-1}(i\omega_n) \right|^2,
\end{align}
where $W_n$ is a weight function. Popular choices for the weight function are $W_n = 1$ , $W_n=1/\omega_n$ and $W_n=1/\omega_n^2$. Different weight functions put different emphasis of low/higher Matsubara frequencies \cite{senechal_bath_2010}. Throughout this work, we have chosen $\beta=40$ and $n_\text{max}=1000$. This method only provides the effective parameters $\tilde \varepsilon_1$ and $\tilde V$. However, in order to calculate the functional $\tilde \Phi [\rho_{\tilde H}]$ an optimal unitary transformation in above sense is calculated and the $h^\mathcal{R}_{ikk'}$ are found by a projection of a Hartree-Fock solution onto the basis states of $\mathcal{R}$. We assume that the effective energy of the impurity site is the same as in the original model ($\tilde \varepsilon_d = \varepsilon_d$).

\bibliographystyle{apsrev4-1}
\bibliography{BibliogrGrafeno}

\end{document}